\documentclass[12pt]{article}
\usepackage[english]{babel}
\usepackage{epsfig}
\usepackage{lscape}
\newcommand{\be}{\begin{equation}}
\newcommand{\ee}{\end{equation}}
\newcommand{\lb}[1]{\label{#1}}
\newcommand{\bi}[1]{\bibitem{#1}}
\begin{document}
\begin{large}
\begin{center}
{\Large \bf The problems of determing the Breit--Wigner parameters of nucleon 
resonance $S_{11}(1535)$}\\[5mm]

E.V.Balandina, E.M.Leikin and N.P.Yudin\\[3mm]
Scobeltsyn Institute of Nuclear Physics,\\
Moscow State University,\\ Russia\\
\end{center}
\section{Introduction}

This paper completes the  phenomenological analysis of the set of 
experimental data on $\eta$-meson
photoproduction  on protons. The main goal of the analysis was to receive
the reliable estimates of some fundamental parameters, first of all the
multipole amplitudes of the process.
The most direct and statistically justified way is to divide the whole analysis
into two stages. The first one is the partial waves analysis of the
experimentally measured observables, for example, angular distributions. The
contribution of different partial waves can be obtained by linear regression
procedure based on linear nonparametric model. Expanding the observables into
series, in this case, over the orthogonal Legendre polinomials we can determine
the terms significantly different from zero and reveal partial waves
participating in the process. Thus, the first
stage of analysis give us the number of the linear regression coefficients that
provide   accordingly statistics the best
description of the experimental observables.

\section{Results of partial wave analisys} 

The result of the partial wave analysis of the angular distributions measured by
three experimental groups \cite{krus1,graal1} and \cite{soez} was presented in
\cite{bal3eng}. The main conclusions from this analysis are:\\
1. One confirms  the known fact that in  energy region from
threshold to about 1GeV $\eta$-mesons are produced mainly in $s$-state and 
the  process has the resonance
character.\\
2. The contributions of  higher partial waves appear only as $sp$- and
$sd$-interference  and their contributions  are small
compared to $s$-wave. Unfortunately the results of three experiments in this respect
qualitatively contradict with each other.

We made also the partial wave analysis of angular distribution measured in gamma
ray region 0.795 -- 1.925 GeV \cite{jlab1}. In this case like the lower energy 
region the linear model with first three terms in the expansion of the
diffential cross section of the process over  Legendre polinomials was proved to
be also statistically justified. The energy dependence of corresponding
regression coefficients $a_0,\ a_1$ and $a_2$ is shown of Fig.1 \cite{bal4}.
These figures confirm the dominant role of the $s$-wave and presence the $p$-
and $d$-partial waves by order of magnitude smaller. There is no evidence of
resonsnce contribution besides $S_{11}(1535)$, firmly established in energy
region between 1 and 2 GeV, and namely $S_{11}(1650)$, $D_{15}(1675)$,
$F_{15}(1680)$ and $P_{13}(1720)$ \cite{RPP}.

 Smooth energy behaviour of the regression
coefficients $a_0, \ a_1$ and $a_2$ seems to indicate that in this energy region
the principal role are playing the nonresonans waves.

\begin{figure}
\parbox{15cm}{
\epsfxsize=15cm
\epsfysize=6cm
%\begin{center}
\hspace*{3mm}
\epsfbox{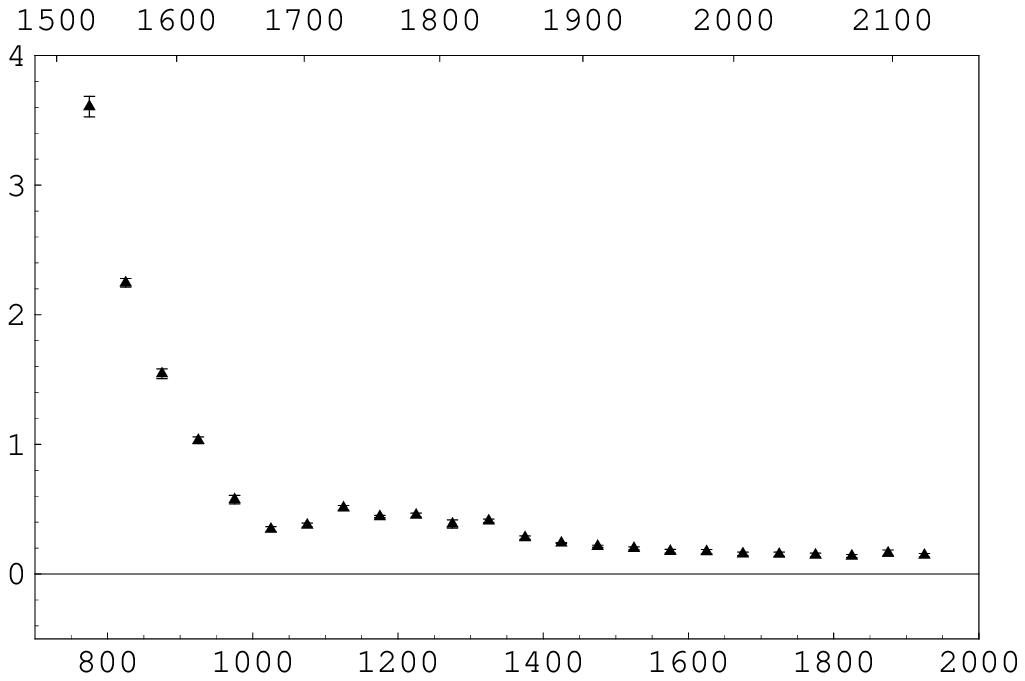}
%\end{center}
\begin{picture}(0.3,0.3)
\put(7.,7.){\footnotesize{$W$, MeV}}
\end{picture}
\begin{picture}(0.3,0.3)
\put(10.5,5.7){\normalsize{$a_0$, $\mu b/sr$}}
\end{picture}
}
\parbox{15cm}{
\epsfxsize=15cm
\epsfysize=6cm
%\begin{center}
\epsfbox{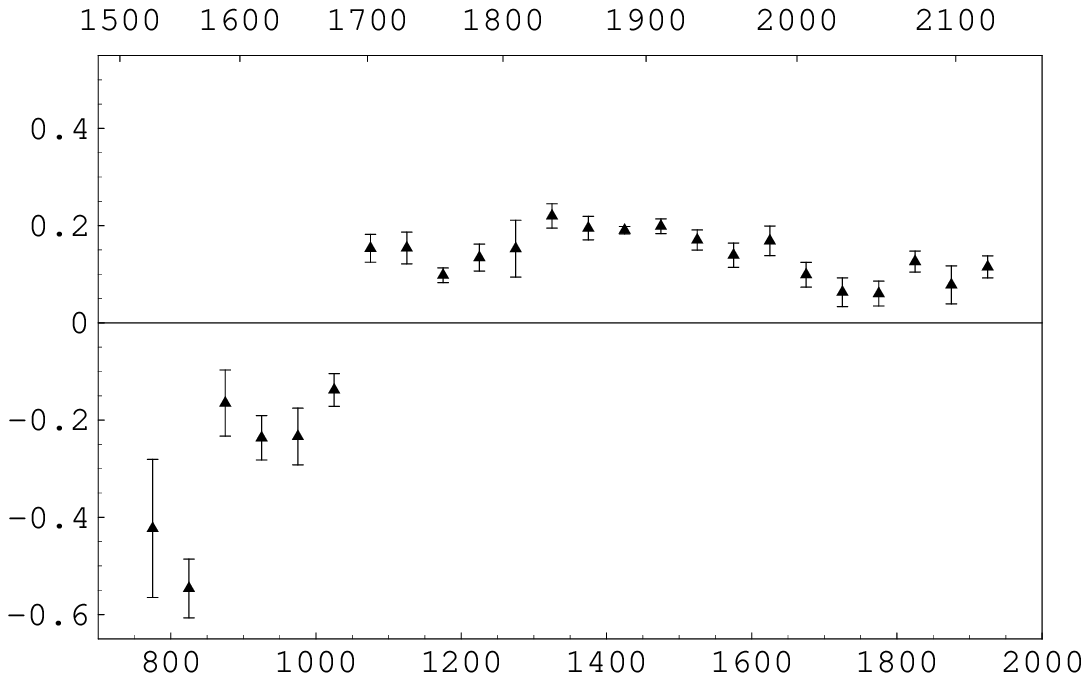}
%\end{center}
\begin{picture}(0.3,0.3)
\put(11.,5.7){\normalsize{$a_1$, $\mu b/sr$}}
\end{picture}
}
\parbox{15cm}{
\epsfxsize=15cm
\epsfysize=6cm
%\begin{center}
\epsfbox{jl1.eps}
%\end{center}
\begin{picture}(0.3,0.3)
\put(11.,5.7){\normalsize{$a_2$,$\mu b/sr$}}
\end{picture}
\begin{picture}(0.3,0.3)
\put(7.,0.2){\footnotesize{$E_\gamma$, МэВ}}
\end{picture}
}
%\vspace*{-1cm}
\caption{The energy dependence of the regression coefficients $a_0,\ a_1,\ a_2$
($W$ -- total energy in CM system) from \cite{jlab1}}
\label{fig1}
\end{figure}

\newpage
\section{ Breit--Wigner characteristics of $S_{11}(1535)$-resonance and
$E_{0^+}$-amplitude of $\eta$-meson photoproduction on protons}

The results  of the  partial wave analysis   of the 
photoproduction of $\eta$-mesons on protons in the energy region from threshold
to about 2 GeV permit the significant simplification of the second part of the
phenomenological analysis that is determination of the multipole amplitudes and
parameters of  $S_{11}(1535)$ resonance. The statistically justified descripton
of experimental observables with three terms of the expansion over  Legendre polinomials
means $s$-wave dominance. So in the expression of $a_0,\ a_1$ and $a_2$ in terms
of multipole amplitudes presented in \cite{bal3eng} we can  keep only the
terms with $E_{0^+}.$  The main contribution of 
$s$-wave contains $a_0$:
\be
a_0(W)=|E_{0+}|^2 \,,
\ee
$W$ -- total energy in CM-system.

Assuming the Breit--Wigner behaviour of the  electric dipole amplitude we can 
use this fact as the parametric statistical model for 
description of  regression coefficient $a_0(W)$ or  the total cross section
$(k/q)\sigma (W)=4\pi a_0(W),$ where $k$ and $q$ -- CM-momenta of gamma quantum and
$\eta$-meson correspondingly. The relevant formulae that take into account the
energy dependence of the resonance width are \cite{krus1}:
\be
\lb{2}
{k\over q}\sigma (W)=4\pi a_0(W)=
{\sigma (W_R)W^2_R\Gamma^2_R\over (W^2_R-W^2)^2+W^2_R \Gamma ^2(W)}\,,
\ee

\be
\lb{3}
\Gamma (W)=\Gamma_R(b_\eta {q_\eta\over q_{\eta R}}+
b_\pi {q_\pi\over q_{\pi R}}+b_{\pi\pi})\,.
\ee
where $\Gamma$ is total width of the resonance, 
$b_\eta,\ b_\pi,\ b_{\pi\pi}$ are the probabilities of  decay to corresponding 
channels,  $q_\eta,\ q_\pi$     
are the momenta $\eta$ and $\pi$,  index R means the value at resonance. 
There were three fitted parameters: $\sigma(W_R),\ W_R$ and $\Gamma_R,$ which
were used to calculate also helisity amplitude: 
\be
\lb{4}
A^2_{1/2}={W_R\over 2m_p}{\Gamma_R\over b_\eta} \sigma(W_R)\,,
\ee
($m_p$ -- proton mass) and  parameter $\xi$ 
that doesn't depend on  $b_\eta$
\be
\lb{5}
\xi ={1\over 2} \sqrt{{k\over q} \sigma(W_R)}\,.
\ee

The values of $b_\eta,\ b_\pi,\ b_{\pi\pi}$ were fixed. The results are presented
in tables 1-4.
Table 1 presents   the estimates received by fitting both $4\pi a_0(W)$ and
$(k/q)\sigma (W)$ which agree quite well. But there is the large inconsistency
between total width of  $S_{11}(1535)$ resonance obtained from the experimental 
data of paper \cite{krus1} and \cite{graal1} and experimental data of papers 
\cite{soez} and \cite{jlab1}. 

\begin{table}[h]
\caption{\lb{tab1} Parameters of resonance $S_{11}(1535)$  from the 
\cite{krus1}}
\begin{center}
\begin{tabular}{|l|l|l|l|}
\hline
$(k/q)(d\sigma/ d\Omega)$&$b_\eta=0.55\,,b_\pi=0.35$&$b_\eta=0.45\,,b_\pi=0.45$&$b_\eta=0.35\,,b_\pi=0.55$\\
\hline
$\sigma(W_R),\ \mu b$ &$33.51\pm 1.5$&$35.88\pm 1.48$&$38.18\pm 1.57$\\
$W_R,$ MeV&$1563.68\pm 9.53$&$1555.32\pm 8.32$&$1547.68\pm 7.86$\\
$\Gamma_R,$ MeV&$270.42\pm 41.18$&$264.02\pm 40.5$&$271.34\pm 46.61$\\
$\xi, 10^{-4}\mbox{ MeV}^{-1}$&$2.08\pm 0.05$&$2.15\pm 0.04$&$2.22\pm 0.05$\\
$A_{1/2}, 10^{-3}\mbox{ GeV}^{-1/2}$&$129.25\pm 6.78$&$142.32\pm 7.46$&$164.28\pm
9.53$\\
$\chi^2/\nu$&$1.2$&$1.21$&$1.24$\\
\hline\hline
$4\pi a_0(W)$&$b_\eta=0.55\,,b_\pi=0.35$&$b_\eta=0.45\,,b_\pi=0.45$&$b_\eta=0.35\,,b_\pi=0.55$\\
\hline
$\sigma(W_R),\ \mu b$ &$34.63\pm 0.82$&$36.92\pm 0.8$&$39.25\pm 0.76$\\
$W_R,$ MeV&$1558.73\pm 4.79$&$1551.32\pm 4.16$&$1543.9\pm 3.57$\\
$\Gamma_R,$ MeV&$244.09\pm 20.32$&$238.41\pm 19.83$&$240.16\pm 20.44$\\
$\xi, 10^{-4}\mbox{ MeV}^{-1}$&$2.11\pm 0.02$&$2.18\pm 0.02$&$2.25\pm 0.02$\\
$A_{1/2}, 10^{-3}\mbox{ GeV}^{-1/2}$&$122.96\pm 3.51$&$135.31\pm 3.83$&$154.43\pm
4.42$\\
$\chi^2/\nu$&$0.52$&$0.52$&$0.49$\\
\hline
\end{tabular}
\end{center}
\end{table}

\begin{table}[h!]
\caption{\lb{tab2} Parameters of resonance $S_{11}(1535)$ from \cite{graal1}}
\begin{center}
\begin{tabular}{|l|l|l|l|}
\hline
$4\pi a_0(W)$&$b_\eta=0.55\,,b_\pi=0.35$&$b_\eta=0.45\,,b_\pi=0.45$&$b_\eta=0.35\,,b_\pi=0.55$\\
\hline
$\sigma(W_R),\mu b$ &$41.06\pm 1.24$&$42.89\pm 1.28$&$44.79\pm 1.35$\\
$W_R,$ MeV&$1537.33\pm 1.47$&$1533.53\pm 1.54$&$1529.4\pm 1.69$\\
$\Gamma_R,$ MeV&$139.7\pm 5.76$&$137.08\pm 5.49$&$134.99\pm 5.3$\\
$\xi, 10^{-4}\mbox{ MeV}^{-1}$&$2.3\pm 0.03$&$2.35\pm 0.04$&$2.4\pm 0.04$\\
$A_{1/2}, 10^{-3}\mbox{ GeV}^{-1/2}$&$93.43\pm 1.54$&$102.71\pm 1.67$&$115.62\pm
1.9$\\
$\chi^2/\nu$&$15.27$&$15.29$&$15.54$\\
\hline
\end{tabular}
\end{center}
\end{table}

\begin{table}[h!]
\caption{\lb{tab3} Parameters of resonance $S_{11}(1535)$  from \cite{soez}}
%\vspace*{-0.5cm}
\begin{center}
\begin{tabular}{|l|l|l|l|}
\hline
$4\pi a_0(W)$&$b_\eta=0.55\,,b_\pi=0.35$&$b_\eta=0.45\,,b_\pi=0.45$&$b_\eta=0.35\,,b_\pi=0.55$\\
\hline
$\sigma(W_R),\mu b$ &$29.83\pm 0.76$&$31.16\pm 0.71$&$32.51\pm 0.66$\\
$W_R,$ MeV&$1559.18\pm 2.83$&$1554.\pm 2.46$&$1548.47\pm 2.13$\\
$\Gamma_R,$ MeV&$199.61\pm 13.86$&$196.17\pm 13.43$&$194.25\pm 13.25$\\
$\xi, 10^{-4}\mbox{ МэВ}^{-1}$&$1.96\pm 0.02$&$2.\pm 0.02$&$2.05\pm 0.02$\\
$A_{1/2}, 10^{-3}\mbox{ GeV}^{-1/2}$&$103.35\pm 2.45$&$113.83\pm
2.63$&$128.63\pm2.92$\\
$\chi^2/\nu$&$2.3$&$2.3$&$2.31$\\
\hline
\end{tabular}
\end{center}
\end{table}

\begin{table}[h!]
\caption{\lb{tab4} Parameters of resonance $S_{11}(1535)$ from \cite{jlab1}}
%\vspace*{-0.7cm}
\begin{center}
\begin{tabular}{|l|l|l|l|}
\hline
$4\pi a_0(W)$&$b_\eta=0.55\,,b_\pi=0.35$&$b_\eta=0.45\,,b_\pi=0.45$&$b_\eta=0.35\,,b_\pi=0.55$\\
\hline
$\sigma(W_R),\mu b$&$43.18\pm 9.13$&$46.15\pm 11.43$&$49.25\pm 14.19$\\
$W_R,$ MeV&$1530.63\pm 12.13$&$1525.6\pm 13.74$&$1520.34\pm 15.46$\\
$\Gamma_R,$ MeV&$152.4\pm 18.7$&$147.94\pm 16.4$&$143.97\pm 14.53$\\
$\xi, 10^{-4}\mbox{ МэВ}^{-1}$&$2.36\pm 0.25$&$2.44\pm 0.3$&$2.52\pm 0.36$\\
$A_{1/2}, 10^{-3}\mbox{ GeV}^{-1/2}$&$96.86\pm 8.32$&$106.03\pm 10.47$&$118.47\pm
13.74$\\
$\chi^2/\nu$&$11.22$&$11.22$&$11.25$\\
\hline
\end{tabular}
\end{center}
\end{table}

It shoud be noted that the later estimates agree
with values of $\Gamma$ obtained from hadron processes and thereby remove the
disagreement existed in values of $\Gamma$ estimated from photo- and
hadroproduction of $\eta$-mesons \cite{RPP}. The differences between total width
and other parameters resulting from experimental data of various experiments are
illustrating by Breit-Wigner curves fitted with  $b_\eta =0.55,\ b_\pi=0.35$ 
on the figures 2-4.

\begin{figure}
\parbox{7cm}{
\epsfxsize=7cm
\epsfysize=7cm
\begin{center}
\epsfbox{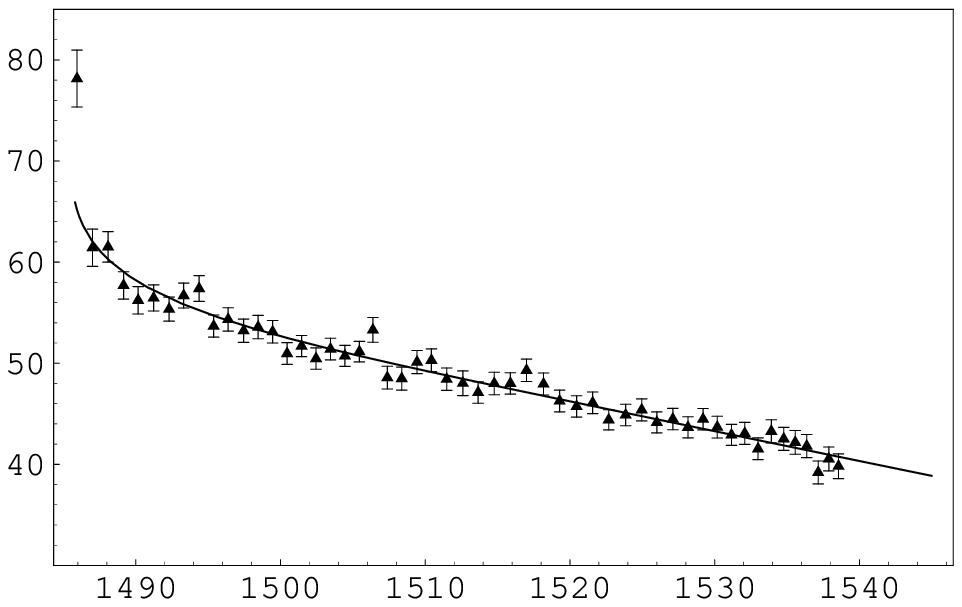}
\begin{picture}(0.3,0.3)
\put(-1.,8.){\footnotesize{${k\over q} \sigma$, $\mu b$}}
\end{picture}
\begin{picture}(0.3,0.3)
\put(0.,0.2){\footnotesize{$W$, MeV}}
\end{picture}
\end{center}
}
\hspace{1cm}
\parbox{7cm}{
\epsfxsize=7cm
\epsfysize=7cm
\begin{center}
\epsfbox{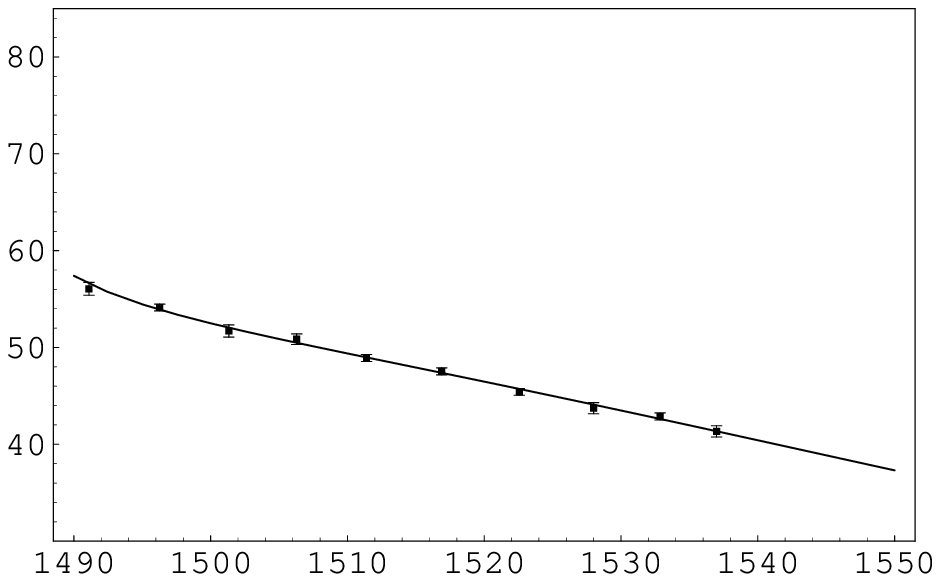}
\begin{picture}(0.3,0.3)
\put(-1.,8.){\footnotesize{$4\pi a_0(W)$, $\mu b$}}
\end{picture}
\begin{picture}(0.3,0.3)
\put(0.,0.2){\footnotesize{$W$, MeV}}
\end{picture}
\end{center}
}
\vspace*{-0.3cm}
\caption{The comparison of experimental  points from \cite{krus1} 
with parametric model (\ref{2}), (\ref{3}) with
$b_\eta =0.55,\ b_\pi=0.35$.}
\label{fig2}
\end{figure}

\begin{figure}
\epsfxsize=8cm
\epsfysize=8cm
\begin{center}
\epsfbox{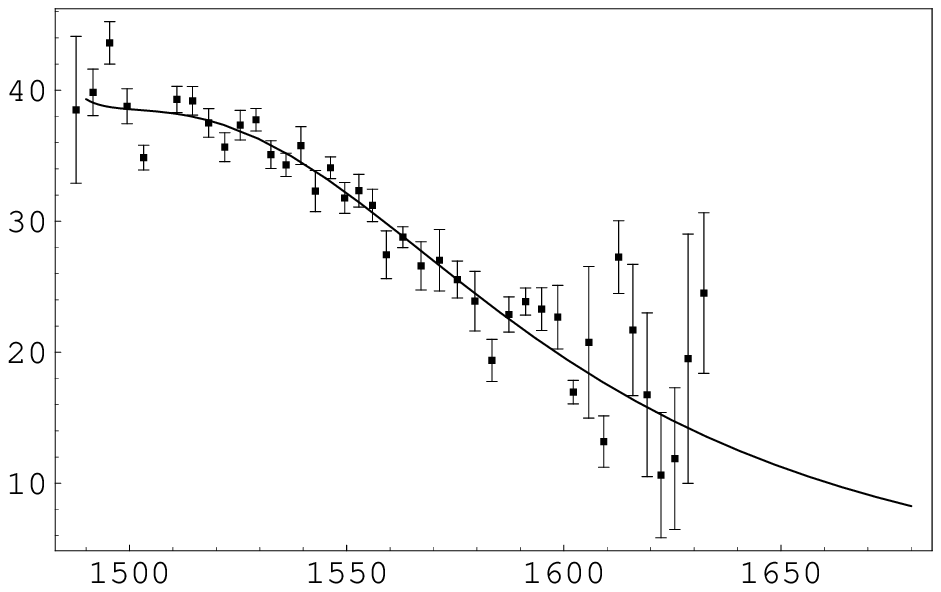}
\begin{picture}(0.3,0.3)
\put(-5.,8.3){\footnotesize{$4\pi a_0(W)$, $\mu b$}}
\end{picture}
\begin{picture}(0.3,0.3)
\put(-5.,-0.3){\footnotesize{$W$, MeV}}
\end{picture}
\end{center}
%\vspace*{-0.5cm}
\caption{The comparison of  experimental  points from \cite{soez} 
with parametric model (\ref{2}), (\ref{3}) with 
$b_\eta =0.55,\ b_\pi=0.35$.}
\label{fig3}
\end{figure}

\begin{figure}
\parbox{7cm}{
\epsfxsize=7cm
\epsfysize=7cm
\begin{center}
\epsfbox{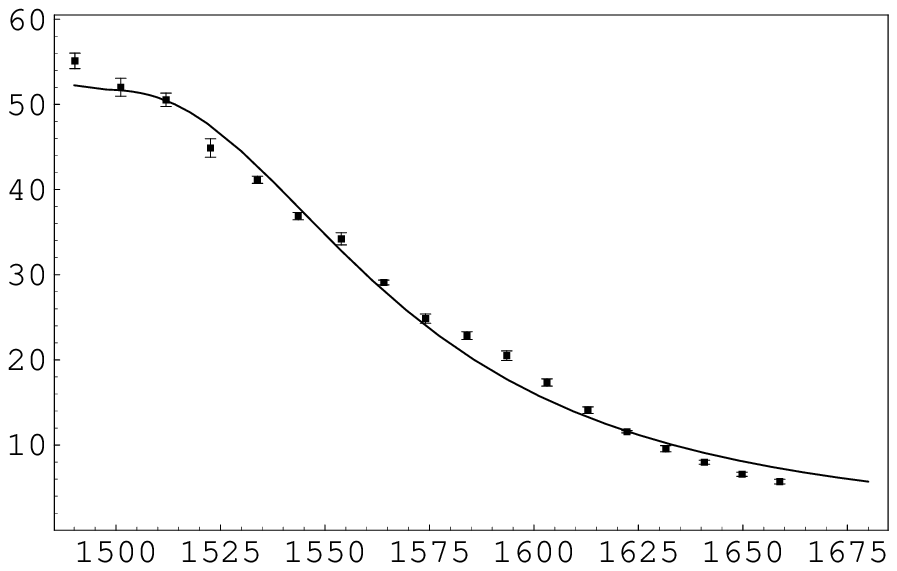}
\begin{picture}(0.3,0.3)
\put(-0.7,7.8){\footnotesize{$4\pi a_0(W)$, $\mu b$}}
\end{picture}
\begin{picture}(0.3,0.3)
\put(0.,0.5){\footnotesize{$W$, MeV}}
\end{picture}
\begin{picture}(0.3,0.3)
\put(2.,7.){\normalsize{(a)}}
\end{picture}
\end{center}}
\hspace{1cm}
\parbox{7cm}{
\epsfxsize=7cm
\epsfysize=7cm
\begin{center}
\epsfbox{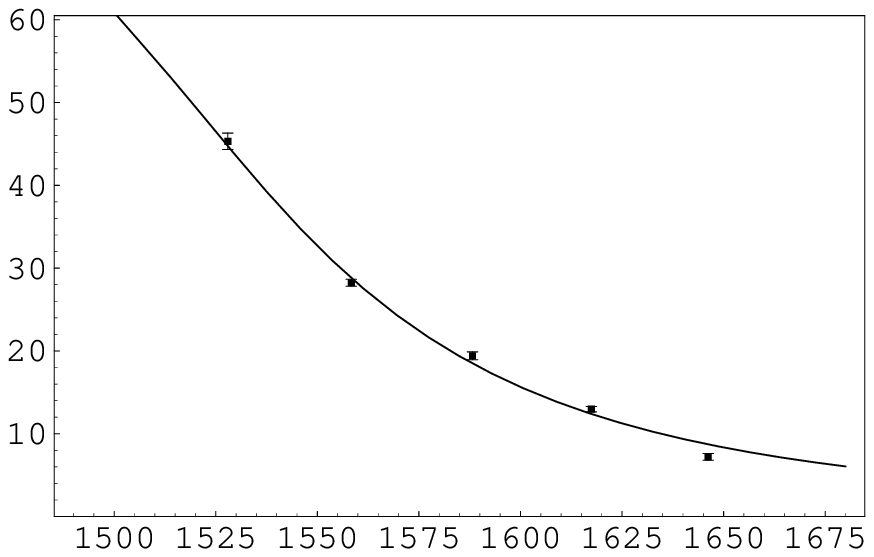}
\begin{picture}(0.3,0.3)
\put(-0.7,7.8){\footnotesize{$4\pi a_0(W)$, $\mu b$}}
\end{picture}
\begin{picture}(0.3,0.3)
\put(0.,0.5){\footnotesize{$W$, MeV}}
\end{picture}
\begin{picture}(0.3,0.3)
\put(2.,7.){\normalsize{(b)}}
\end{picture}
\end{center}}
\vspace*{-1cm}
\caption{The comparison of experimental  points from 
\cite{graal1}\ (a) and \cite{jlab1} (b) 
with parametric model (\ref{2}), (\ref{3}) with  
$b_\eta =0.55,\ b_\pi=0.35$.}
\label{fig4}
\end{figure}

The large values of  criterion $\chi^2/\nu$ per degree of freedom in tables 2
and 4 is due to the considerable deviation of the experimental values relative the
fitted curve which significantly exceed the experimental errors. This means that
the deviations of points from the Breit--Wigner curve do not follow the normal
distribution with zero expectation. In this case the goodness of fit can be
estimated by the value of noncentral criterion 
$\chi^2_{\mbox{\small nc}}$ with 
noncentrality parameters $\delta ^2$  equal to the sum of  expectation squared 
of individual  variables \cite{8}. As an example  in Table 5 the differences 
$\Delta$ between the 
 values $4\pi a_0(W)$
obtained as a result of partial waves analysis of experimental data 
\cite{graal1} and 
the Breit--Wigner curve for the case $b_\eta=0.55,\ b_\pi=0.35$ are shown. 
\begin{table}
\caption{ 
\label{tab5}
 The differences  $\Delta$ between experimental and fitted values of 
$4\pi a_0(W)$ from \cite{graal1}.}
\begin{center}
\begin{tabular}{|c|c||c|c||c|c|}
\hline
$W$, MeV&$\Delta$&$W$, MeV&$\Delta$&$W$, MeV&$\Delta$\\
\hline
1490.16&2.9&1553.9&1.41&1612.92&0.83\\
1501.14&0.36&1564.08&0.98&1622.26&$-0.06$\\
1511.98&0.09&1574.06&0.82&1631.55&$-0.67$\\
1522.61&$-2.6$&1583.98&2.3&1640.78&$-1.09$\\
1533.85&$-1.59$&1593.49&2.78&1649.79&$-1.53$\\
1543.61&$-1.06$&1603.12&2.03&1658.81&$-1.55$\\
\hline
\end{tabular}
\end{center}
\end{table} 

These differences were used to estimate the value of 
noncentrality  parameter   $\delta^2=46.8.$ 
The  expectation of noncentral value 
$\chi^2_{\mbox{\small nc}}$ per degree of freedom turns out to be 1.05 
supporting   
statistical 
reliability of the result presented in  tables 2 and 4. At the same time this
item may be considered as some indication of possible biases in experimental
data.

Description of energy dependence  $4\pi a_0(W)$   by  parametric 
model  based on the Breit--Wigner 
formula allows to find the real and imaginary parts  and 
also the phase of electric dipole amplitude $E_{0+}$   responsible for 
production of $S_{11}(1535)$- 
resonance. 

As an example on fig.5 the data obtained from the  energy 
dependence   $4\pi a_0(W)$  from paper \cite{graal1} are  shown. 

\begin{figure}[h!]
\vskip 1cm
\parbox{5.cm}{
\vspace*{-2mm}
\epsfxsize=5.cm
\epsfysize=7.15cm
\epsfbox{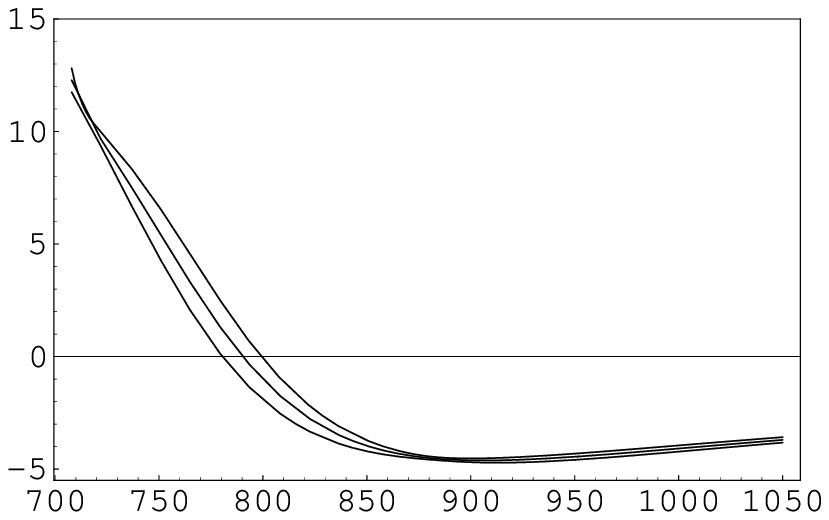}
\begin{picture}(0.3,0.3)
\put(-0.,8.2){\footnotesize{Re$E_0^+, 10^{-3}/m_\pi$}}
\end{picture}
\begin{picture}(0.3,0.3)
\put(1.5,0.5){\footnotesize{$E_\gamma$, MeV}}
\end{picture}
\begin{picture}(0.3,0.3)
\put(3.,6.8){\normalsize{(a)}}
\end{picture}
}
\parbox{5.cm}{
\epsfxsize=5.cm
\epsfysize=7cm
\epsfbox{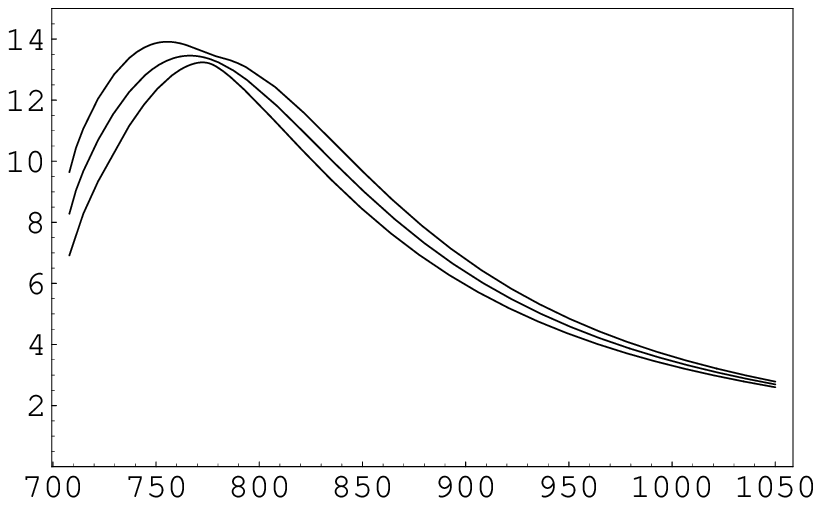}
\begin{picture}(0.3,0.3)
\put(-0.,8.2){\footnotesize{Im$E_0^+, 10^{-3}/m_\pi$}}
\end{picture}
\begin{picture}(0.3,0.3)
\put(1.5,0.5){\footnotesize{$E_\gamma$, MeV}}
\end{picture}
\begin{picture}(0.3,0.3)
\put(3.,6.9){\normalsize{(b)}}
\end{picture}
}
\parbox{5.cm}{
\vspace*{-.5mm}
\epsfxsize=5.cm
\epsfysize=7.05cm
\epsfbox{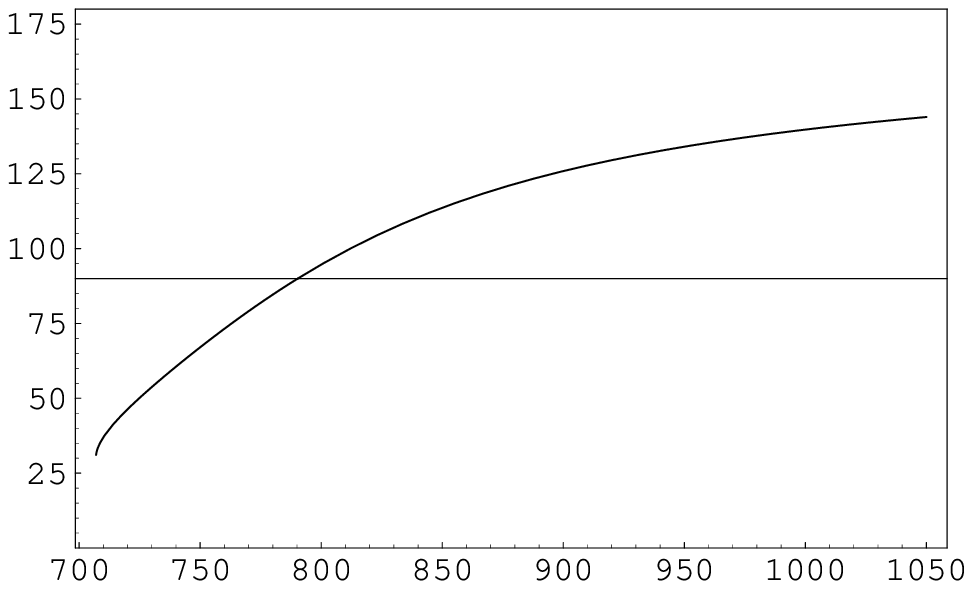}
\begin{picture}(0.3,0.3)
\put(-0.,8.2){\normalsize{$\varphi$}}
\end{picture}
\begin{picture}(0.3,0.3)
\put(1.6,0.6){\footnotesize{$E_\gamma$, MeV}}
\end{picture}
\begin{picture}(0.3,0.3)
\put(3.,6.9){\normalsize{(c)}}
\end{picture}
}
\caption{The energy dependence of the real     
(a), imaginary (b) and phase $\varphi$ (c) of  the multipole amlitude $E_{0+}$ 
from \cite{graal1}.}
\label{fig5}
\end{figure}

Upper and lower curves were calculated  with  error matrix of parameters 
$W_R,\ \sigma(W_R)$ и $\Gamma_R$ calculated with  $b_\eta=0.55,\
b_\pi=0.35$ from table 2:
$$
\left(
\begin{array}{ccc}
1.54  &-1.23 &-4.71\\
-1.23 &2.16  &1.84\\
-4.71 &1.84  &33.20\\
\end{array}
\right)
$$

Real and imaginary  parts of $E_{0+}$  transforms the  regression
coefficients  containing   $E_{0+}$ into linear equations. 
Unfortunately the regression 
coefficients  of polarization observables have significant errors and
are at different 
energies \cite{bal3eng}. This means that up to now there is no possibility
 to find the reliable estimates of higher multipole. 
In order to go further  and find reliable estimates 
of $p$-, $d$- and higher multipole  it is necessary to have more precise and more
detailed experimental data on polarization observables including  
angular distributions of  recoiled protons polarization.

\section{ Conclusion}
The partial wave analysis of observables  of eta photoproduction on protons 
made in \cite{bal3eng}  confirms the dominating role of the s-wave in energy 
region from threshold up to about 2 GeV and demonstrates its resonance behaving 
 corresponding to production of nucleon resonance $S_{11}(1535).$ The 
experimental data on differential cross sections don't  show  the presence of 
the partial waves with $l>2.$ In all cases the statistically reliable description of 
experimental data was achieved with keeping only three terms in expansion of 
differential cross section over Legendre polynomials.
Contributions of $p$- and $d$-waves appear only through their interference with the 
resonance $s$-wave  are small and the 
results of different experiments contradict with each other. 
 Thus in  the energy region up 
to 2 GeV the experimental data doesn't show appearence of large number of 
firmly established resonances \cite{RPP}.

The separation of $s$-wave contribution has allowed 
to find the energy dependence of   amplitude $E_{0+}$  and fundamental 
characteristics of 
$S_{11}(1535)$ resonance. Unfortunately the data of various experiments  in
this case are also incompatable.

\end{large}
\end{document}